\documentstyle[twoside,fleqn,espcrc2]{article}
\input epsf

\newcommand{\AmS}{{\protect\the\textfont2
  A\kern-.1667em\lower.5ex\hbox{M}\kern-.125emS}}

\title{Analytical Results for Abelian Projection}

\author{Michael C. Ogilivie\address{Department of Physics,
        Washington University, \\ 
        St. Louis, MO 63130 USA}%
        \thanks{I thank the U.S. Department of Energy
	for financial support under
	grant number DE-FG02-91-ER40628.}
}
       
\begin{document}

\begin{abstract}
Analytic methods for Abelian projection are developed,
and a number of results related to string tension measurements are
obtained.
It is proven that even without gauge fixing,
Abelian projection yields string tensions of the underlying non-Abelian
theory.
Strong arguments are given for similar results in the case
where gauge fixing is employed.
The subgroup used for projection need only contain the center
of the gauge group,
and need not be Abelian.
While gauge fixing is shown to be in principle
unnecessary for the success of Abelian projection,
it is computationally advantageous for the same reasons that improved
operators, e.g., the use of fat links, are advantageous
in Wilson loop measurements. 
\end{abstract}

\maketitle

\section{Introduction}

Although Abelian projection has a compelling theoretical basis
\cite{'t Hooft}\cite{Mandelstam}, and has notable sucesses
\cite{Suzuki} in lattice simulations, there
are several fundamental questions which
remain unresolved.\cite{Greensite}
The most basic issue is the possibility that
the successes of abelian projection are not
consequences of our insight into the non-abelian dynamics,
but simply reflect general field-theoretic principles.
A closely related, more practical issue is the
choice of the correct or best subgroup to use for projection.
As discussed below, gauge invariance is the key 
physical principle responsible for the success of Abelian projection;
furthermore, any subgroup, not necessarily Abelian, will work as long
as it contains the center of the gauge group.
Details are given in reference \cite{mco-projection}.

\section{Abelian Projection in Practice}

\vspace{1pt}The standard approach to Abelian projection is a three step
process. The gauge fields $g_l$ are associated with links of the lattice, and
take on values in a compact Lie group $G$. An ensemble of lattice gauge
field configurations is generated using standard Monte Carlo methods.
Each field configuration in the $G$-ensemble is placed in a particular
gauge by site-based gauge transformations $\phi_s$.
The gauge-fixing condition is
chosen to preserve gauge invariance for some subgroup $%
H $ of $G$. A typical gauge-fixing procedure for Abelian projection
is to maximize
\begin{eqnarray}
S_{gf}=\lambda \sum_{l}Tr\,\left[ g_{l}\,Mg_{l}^{+}M\right]
\end{eqnarray}
where $M$ is a traceless, Hermitian matrix that commutes with
every element of the subgroup $H$ and the sum is taken over all links.
From this ensemble of gauge-fixed field configurations, another ensemble of
gauge fields is generated, with the fields
taking on values in the subgroup $H$. This is obtained by maximizing 
\begin{eqnarray}
S_{proj}\left[ g,h\right] =\sum_{l}\left[ \frac{p}{2N}Tr\left(
g_{l}^{+}h_{l}+h_{l}^{+}g_{l}\right) \right]
\end{eqnarray}
where $h_{l}\in H$.

For analytical purposes, it is necessary to generalize the
projection procedure,
so that a single configuration of $g$-fields will be associated with an
ensemble of configurations of both $h$-fields and $\phi$-fields.
The original fields $g_{\mu }(x)$ are replaced in the gauge-fixing function
and the projection function by
$\widetilde{g}_{\mu }(x)=\phi (x)g_{\mu }(x)\phi ^{+}(x+\widehat{\mu })$.
For each $g$-field configuration, an ensemble of $\phi$-fields
will be generated, weighted by $S_{gf}$.
For each $\tilde{g}$-field configuration, an ensemble of
$h$-fields will be generated, weighted by $S_{proj}$. 
The parameters $\lambda$ and $p$ control the width of the two
distributions, and the original scheme is formally regained in the
limit $\lambda$, $p \rightarrow \infty $. 
The distribution of the ensemble of $g$-fields is independent
of $\lambda$ and $p$.
The field $\phi$ is
quenched relative to $g$, and $h$ is quenched relative to
both $\phi$ and $g$. This quenching is necessary to preserve
the normal lattice definition of gauge-invariant quantities,
but introduces complications reminiscent of spin glasses 
into expressions for expectation values.

\section{Projection without Gauge Fixing}

The simplest and most analytically tractable case is
projection without gauge fixing, which is achieved by setting
$\lambda = 0$.
In this case, the integration over the $\phi$ fields
takes any configuration overs its entire gauge orbit,
thus serving to enforce gauge invariance.

Consider the expectation value of a Wilson loop $W$ 
with no self-intersections
in a representation $\beta$ of $H$, denoted by
$ \left\langle \widetilde{\chi }^{\beta }(W)\right\rangle = 
\left\langle \widetilde{\chi }^{\beta }(h_{1}..h_{n})\right\rangle$
where $\widetilde{\chi }^{\beta }$ stands for the appropriate
group character.
In calculating the expectation value,
the weight function for projecting each link can be expanded in the
characters of the group $G$, 
\begin{eqnarray}
\lefteqn{ \exp \left[ \frac{p}{2N}Tr\left( g^{+}h+h^{+}g\right) \right]
\nonumber }\\
&& = \sum_{\alpha} 
d_{\alpha }c_{\alpha }\left( p\right) \chi _{\alpha }\left( h^{+}g\right) .
\end{eqnarray}
The principal difficulty in the evaluation of the expected
value lies in the treatment of factors resulting from the quenching
of $\phi$. To lowest order in the character expansion,
systematic application of gauge invariance at all sites along the curve $W$
collapses the sum into the simple result 
\begin{eqnarray}
\lefteqn{
\left\langle \widetilde{\chi }^{\beta }(h_{1}..h_{n})\right\rangle
\nonumber } \\
&&=\sum_{\alpha }\left( \frac{c_{\alpha }(p)}{c_{0}(p)}\right)
^{n}\int_{H}(dh)\,\widetilde{\chi }^{\beta }(h)\chi ^{\alpha
}(h^{+}) \nonumber \\
&& \ \ \ \cdot \left\langle \chi ^{\alpha }(g_{1}..g_{n})\right\rangle 
\end{eqnarray}
where the sum is over representations $\alpha$ of $G$.
This formula has obvious
physics content: the Wilson loop as measured in the $\beta $ representation
of $H$ is given as a sum of Wilson loops in the irreducible representations
of $G$, each weighted by the number of times $\beta $ appears in $\alpha $
and by a $p$-dependent factor which contributes to the perimeter dependence.
This result can be turned into rigorous upper and lower bounds on
$ \left\langle \widetilde{\chi }^{\beta }(W)\right\rangle $
with changes only in the perimeter dependence on the right-hand side
of the formula. This leads immediately to 
area law behavior for $H$,
independent of $p$:
\begin{eqnarray}
\tilde{\sigma}_{\beta }=\min_{\alpha }\,\sigma _{a}
\end{eqnarray}
where the minimum is taken over all representations $\alpha $ that have a
non-zero contribution. Center symmetry plays an important role here,
causing many potential terms to vanish.
In the case of $SU(N)$ projected to $Z(N)$, a very direct alternative
proof has been given recently by Ambjorn and Greensite.\cite{Ambjorn}

Consider, as an example, the case of $SU(2)$ projected to $U(1)$.
The string
tension is non-zero for the half-integer representations $j=1/2,\,3/2,\,..$
due to the $Z_{2}$ center symmetry, but not for the integer representations.
A typical result is
\begin{eqnarray}
\tilde{\sigma}_{1/2}=\min_{j=1/2,\,3/2,\,..}\,\sigma _{j} 
\end{eqnarray}
but $\tilde{\sigma}_{1}=0$ because of string-breaking in the adjoint
representation: $\sigma _{1}=0$.

\section{Projection with Gauge Fixing}

Projection combined with gauge fixing ($\lambda \ne 0$ 
is less tractable than the $ \lambda = 0$ case.
However, a strong-coupling expansion in $\lambda$\cite{Fachin},
which is convergent for sufficiently small $\lambda$,
indicates similar results.
The lowest order result can be obtained by expanding
the $\lambda = 0$ result in $p$.
The form of the corrections is determined by gauge invariance. 
To order $\lambda^3$,
the corrections to the area- and perimeter-dependence are determined by
\begin{eqnarray}
\lefteqn{
\left\langle \chi \left( h_{1}..h_{n}\right) \right\rangle  \approx
p^{n}K_{1}\left\langle \chi \left(
g_{1}..g_{n}\right) \right\rangle \nonumber }  \\
&&+p^{n}\lambda ^{3}K_{2} \cdot \nonumber \\ 
&& \sum \left\langle \chi \left( g_{1}g_{A}g_{B}g_{C}..g_{n}\right)
 \chi \left(
g_{2}g_{C}^{+}g_{B}^{+}g_{A}^{+}\right) \right\rangle
\end{eqnarray}
where the summation sign indicates that the decoration
of $g_2$ by the staple $g_A g_B g_C$ 
is to be repeated through the entire loop for all directions
orthogonal to the loop. The constants $K_{1}$ and $K_{2}$ are power series
in $p$ and $\lambda $, beginning at order $1$.
These corrections are shown graphically in
Figure 1. From this expression, we can see that the string tension should
still satisfy 
\begin{eqnarray}
\tilde{\sigma}_{\beta }=\min_{\alpha }\,\sigma _{a}
\end{eqnarray}
as in the case of no gauge fixing. The parameters $p$ and $\lambda $ change
the perimeter dependence of the Wilson loop expectation value, but not the
area dependence.
Gauge fixing appears here in a manner quite similar to the use of fat links,
and it is likely that fat links would be numerically advantageous when
projection is performed without gauge fixing.

\section{\protect\vspace{1pt}Conclusions}

The success of Abelian projection appears to have its origin in very general
considerations. The key principle is local gauge invariance. Ultimately, it
is Elitzur's theorem\cite{Elitzur}
that ensures that observables constructed from the
projected field can always be rewritten in terms of gauge-invariant
observables of the underlying gauge fields. Abelian dominance is not
necessary. Note that at no point in the arguments given above has space-time
dimensionality been a consideration, a further indication that Abelian
projection does not depend on some particular set of important field
configurations.

There is one possible weak point in the gauge-fixed case: the
standard gauge fixing algorithm corresponds formally to the limit $\lambda
\rightarrow \infty $, but the strong-coupling expansion in $\lambda $ has a
finite radius of convergence. These arguments will fail
for large $\lambda $ if there is, as seems likely, 
a phase transition along some critical line $\lambda
_{c}\left( \beta \right) $, a function of the gauge coupling $\beta $.
However, even if
there is a phase transition in the gauge-variant sector, it may
be that the strong-coupling region and weak-coupling region are in fact
connected because the critical line has an end-point.
In any event, it remains conceptually difficult to claim that
confinement should
be understood differently for large $\lambda $ and small $\lambda $, because
by construction, the underlying ensemble of non-Abelian gauge fields does
not depend on $\lambda $.

\begin{figure}[htb]
\epsfxsize=75mm \epsfbox{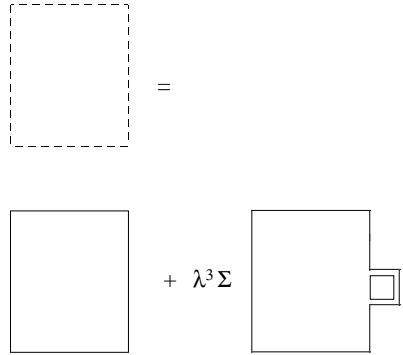}
\vspace{-0.3in}
\caption{Strong-coupling expansion for subgroup $H$ Wilson
loop in terms of $G$ Wilson loops.}
\label{fig:fig1}
\end{figure}

\end{document}